# Revealing the Hidden Patterns: A Comparative Study on Profiling Subpopulations of MOOC Students

**Lei Shi**
*Durham University*
*Durham, UK*                                      *lei.shi@durham.ac.uk*

**Alexandra I. Cristea**
*Durham University*
*Durham, UK*                                *alexandra.i.cristea@durham.ac.uk*

**Armando M. Toda**
*University of São Paulo*
*São Carlos, Brazil*                              *armando.toda@usp.br*

**Wilk Oliveira**
*University of São Paulo*
*São Carlos, Brazil*                              *wilk.oliveira@usp.br*

## Abstract

Massive Open Online Courses (MOOCs) exhibit a remarkable heterogeneity of students. The advent of complex "big data" from MOOC platforms is a challenging yet rewarding opportunity to deeply understand how students are engaged in MOOCs. Past research, looking mainly into overall behavior, may have missed patterns related to student diversity. Using a large dataset from a MOOC offered by FutureLearn, we delve into a new way of investigating hidden patterns through both machine learning and statistical modelling. In this paper, we report on clustering analysis of student activities and comparative analysis on both behavioral patterns and demographical patterns between student subpopulations in the MOOC. Our approach allows for a deeper understanding of how MOOC students behave and achieve. Our findings may be used to design adaptive strategies towards an enhanced MOOC experience.

## 1.   Introduction

In its eighth year (2019), the strong and modern trend of MOOCs (Massive Open Online Courses) has attracted more than 900 universities delivering over 11 thousand courses to more than 100 million students around the world [17]. At a massive scale, online learning demands for different educational approaches designed or adapted to be effective for a large number of students with diverse background [7]. Importantly, the outcomes of a particular learning experience may rely heavily on individual students' differences, including behavioral and demographical aspects [1, 5, 18, 19].

Adaptive educational platforms and student modelling have been proposed, over the past decade, to personalize online learning experience [4, 20, 21] or to author for it [2]. However successful, adaptation has normally been applied to learning environments, where the quantity and diversity of students were relatively small, raising the question of its direct applicability in the "MOOCs context". As such, the challenge is to first reach a deep understanding of MOOCs' massive and diverse students.

To tackle this challenge, past research efforts include those aiming at understanding students' dropouts, completion, motivation and engagement [3, 8, 9, 22]. However, by looking mainly into overall behavior, they may have missed patterns related to student diversity. Thus, in this study, we delve into a new way of investigating hidden patterns through machine learning and statistical modelling. We deploy a large dataset from a MOOC offered by FutureLearn[1], a less explored MOOCs platform comparing its counter parts EdX and Coursera. We particularly focus on two research questions:

*RQ1. How can we find distinct subpopulations of MOOC students?*
*RQ2. Are there behavioral and demographical patterns within subpopulations?*

We answer via a clustering analysis on student activities and comparative analysis on behavioral and demographical patterns between student subpopulations.

## 2. Related Work

Recent studies have provided good examples on how researchers are exploring new methods to understand student behavior in MOOCs. For example, Pursel, et al. [15] examined MOOC students' demographical data, intended behaviors and course interactions to understand variables that are indicative of MOOC completion. They analyzed data from both pre-course survey and system logs from a specific MOOC and identified a list of motivations and behaviors that could influence MOOC completion. The results provided insights into several variables – such as prior degree attainment and course interaction data – that showed relationships to MOOC completion. However, the authors considered the student body as a whole, unlike in our current study, where we partition students, revealing hidden within-group behavior patterns.

Cristea, et al. [6] conducted a data-intensive analysis on a large-scale data collection of 5 MOOCs spread over 21 runs, focusing on temporal quiz solving patterns. In particular, the analyses focused on rates of quiz questions attempted, correctly answered and incorrectly answered, as well as how these rates changed over the course of a term, for both students who completed and who did not complete the course. The result suggested that the completion was significantly correlated to quiz solving behaviors. Whilst the authors only used quiz solving patterns to compare subpopulations of the students, *i.e.* completers and non-completers, in this study, besides quiz behavior we also considered how students visited learning materials and commented on the discussion board.

Rieber [16] investigated students' behavior and patterns of participation in MOOCs. The study adopted a descriptive research design involving survey, quiz and participation data, using various statistical models such as Pearson product-moment correlation coefficient, correlated-samples t-test, and one-way ANOVA, similar to our study. Their results suggested that even highly structured, instructional MOOCs could offer flexible learning environments for students with varied goals and needs. However, their study only used statistical modelling to provide simple descriptive analysis on the whole student cohort; whilst we used not only statistical modelling, but also machine learning techniques, *i.e.* k-means, to cluster students into subpopulations and compared them for in-depth analysis.

Van den Beemt, et al. [23] explored the relation between MOOC students' learning behavior and learning progress, in order to gain insight into how passing and failing students distribute their activities differently along the course. Aggregated counts of activities and specific course items, as well as the order of activities, were examined with techniques including cluster analysis. Four student clusters were identified and compared with each other. However, the variables were limited to video watching scores and quiz submission scores; whilst in our study, we considered the additional and important dimension of social interactions.

Rodrigues, et al. [13] explored the use of cluster analysis, in particular, the hierarchical clustering method (Ward clustering) and non-hierarchical clustering method (k-means) to analyze the engagement behavior characteristics. The result suggested the necessity of meeting the diversity of engagement patterns that allow increasing engagement and fostering a better learning experience. However, the study only considered forum activities; while our study took into account also the important activities of accessing learning materials and taking assessments.

---

[1] www.futurelearn.com

## 3. Dataset and Pre-processing

We used the dataset from the MOOC "Shakespeare and his World" hosted on FutureLearn. In the MOOC, there were 130 *steps*, the basic learning units, across 10 weeks. Each *step* contained an article, a video, or an assessment. Students could leave comments on all *step* pages, except those containing an assessment. Each week included one assessment, and each assessment had 12 questions. Thus, in total, there were 120 questions. Out of the 15,852 initially enrolled students, 1,881 students proactively unenrolled, thus leaving 13,971 students remaining. All of them visited at least one *step* page. The dataset explored in this study included all the data generated by these 13,971 students for the following six variables:

1. **Visits**: the number of distinct *step* pages that a student visited.
2. **Completions**: the number of *steps* a student claimed completion, by clicking the button "Mark as complete" on *step* pages.
3. **Attempts**: the number of questions a student attempted to answer.
4. **Correct answers**: the number of questions a student correctly answered.
5. **Comments**: the number of comments a student posted on *step* pages.
6. **Replies**: the number of replies a student received to all their comments.

As shown in Table 1, in total, the dataset included 511,266 *visits* (M = 36.59, *i.e.* 28.15% *steps* of the course, SD = 45.44), 467,463 *completions* (M = 33.46, *i.e.* 25.73%, SD = 44.87), 1,225,279 *attempts* (M = 87.7, *i.e.* 73.08%, SD = 65.75), 816,756 *correct answers* (M = 58.46, *i.e.* 66.66% of the questions answered, SD = 43.90), 268,797 *comments* (M = 19.24, SD = 45.22), and 123,233 *replies* (M = 8.82, SD = 36.13). Skewness for *visits*, *completions*, *comments* and *replies* above 1 indicates skewed distributions. For *attempts* and *correct answers*, whilst skewness is between -1 and 1, it is greater than three times the standard error for skewness ($.375 > 3 \times .021$), indicating skewed distributions, too. Kurtosis for *comments* and *replies* (positive; kurtosis $\gg$ 1) indicate peaked distributions.

**Table 1.** Descriptive statistics for six selected variables

|  | Mean (SE) | SD | Total | Kurtosis (SE) | Skewness (SE) |
|---|---|---|---|---|---|
| **Visits** | 36.59 (.384) | 45.44 | 511,266 | -.264 (.041) | 1.163 (.021) |
| **Completions** | 33.46 (.380) | 44.87 | 467,463 | -.156 (.041) | 1.206 (.021) |
| **Attempts** | 87.70 (.556) | 65.75 | 1,225,270 | -1.211 (.041) | .375 (.021) |
| **Correct answers** | 58.46 (.371) | 43.90 | 816,756 | -1.651 (.041) | .230 (.021) |
| **Comments** | 19.24 (.383) | 45.22 | 268,797 | 70.731 (.041) | 6.872 (.021) |
| **Replies** | 8.82 (.306) | 36.13 | 123,233 | 336.174 (.041) | 15.669 (.021) |

A further Kolmogorov-Smirnov normality test on these six variables indicated that none was normally distributed ($p < .001$). This shows that students' behavior is clearly not homogenous on any axes (variables) investigated, thus pointing to the clear need of identifying subpopulations with sub-patterns of behavior.

Next, we used correlation to measure the extent to which these variables associate. As the variables were not normally distributed, we used Spearman's rank correlation coefficient. According to [10], the size of correlation (*r* value) has to be less than -.70 or greater than .70, for correlation to be strong. Table 2 displays strong positive ($r > .7$; $p < .001$ [10]) correlations between *visits* and *completions* ($r = .952$, $p < .001$); and between *comments and replies* ($r = .772$, $p < .001$). However, for the rest, *i.e.*, between *visit* and *comments*, etc., correlations were negligible, $r \in \{r \mid -.3 < r < .3\}$. Thus, our selection of clustering algorithm variables was: {*visits*, *attempts*, *comments*}.

Interestingly, this data-driven approach further confirms that these three relatively independent variables represent the fundamental (and, arguably, comprehensive) dimensions of student engagement in FutureLearn MOOCs, *i.e.* learning (*visits*), assessment (*attempts*) and social (*comments*), as previously proposed by us [19].

**Table 2.** Spearman's rho Correlation Coefficient

|  |  | Visits | Completions | Attempts | Comments | Replies |
|---|---|---|---|---|---|---|
| **Visits** | Correlation Coefficient | 1.000 | **.952** | .216 | .101 | .059 |
|  | Significance (2-tailed) |  | *< .001* | *< .001* | *< .001* | *< .001* |
| **Completions** | Correlation Coefficient | **.952** | 1.000 | .213 | .103 | .057 |
|  | Significance (2-tailed) | *< .001* |  | *< .001* | *< .001* | *< .001* |
| **Attempts** | Correlation Coefficient | .216 | .213 | 1.000 | .152 | .091 |
|  | Significance (2-tailed) | *< .001* | *< .001* |  | *< .001* | *< .001* |
| **Comments** | Correlation Coefficient | .101 | .103 | .152 | 1.000 | **.772** |
|  | Significance (2-tailed) | *< .001* | *< .001* | *< .001* |  | *< .001* |
| **Replies** | Correlation Coefficient | .059 | .057 | .091 | **.772** | 1.000 |
|  | Significance (2-tailed) | *< .001* | *< .001* | *< .001* | *< .001* |  |

## 4.   Cluster Analysis

We conducted a cluster analysis using the three variables selected in the pre-processing, *i.e. visits*, *attempts,* and *comments*, as discussed in section 3, to answer the first research question, ***RQ1: How can we find distinct subpopulations of MOOC students?***

In this study, we used the k-means [14] algorithm to cluster students, a well-known unsupervised machine learning algorithm, producing a pre-specified number ($k$) of clusters. To find the optimal $k$, we used the "elbow method" [11], *i.e.* running clustering on a range of values of $k$ (2 ~ 15, in our case) and then calculated the within-group sum of squares (measuring how close cluster members were to its cluster center) and plotted the result on a line chart. The goal was to choose a small value of $k$ that still had a low value of the within-group sum of squares. Before clustering, to make the clustering less sensitive to the scale of the three variables, we standardized them. Fig. 1 shows that, when $k = 7$, the within-group sum of squares dipped down appreciably, so we used it in the k-means algorithm to cluster those 13,971 "active students".

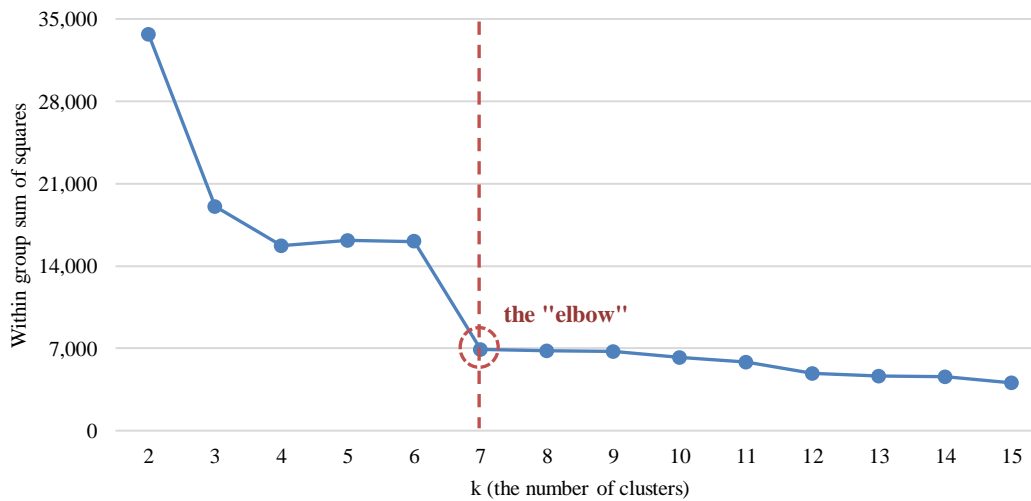

**Fig. 1.** The optimal number of clusters, *i.e.* the $k$ value

When $k = 7$, the convergence of the clustering was achieved in the 14th iteration. Fig. 2 shows the final cluster centers and the size of each cluster. About half of students (7,183; 51.41%) were allocated in Cluster 7, followed by Cluster 1 (3,130; 22.40%) and Cluster 3 (2,797; 20.02%). The most underrepresented cluster was Cluster 2, with only 10 (0.07%) students, followed by Cluster 6 with 66 (0.47%) students. Interestingly, $Z_{comments}$ was the most influential parameter. We can observe from Fig. 2 that students from Cluster 2 (N = 10) and Cluster 6 (N = 66) had much larger positive $Z_{comments}$ values. This indicates that students allocated in Cluster 2 and Cluster 6 could possibly be called amongst the "most social" students, although they were only a very small percentage of the whole population, *i.e.* 0.54%.

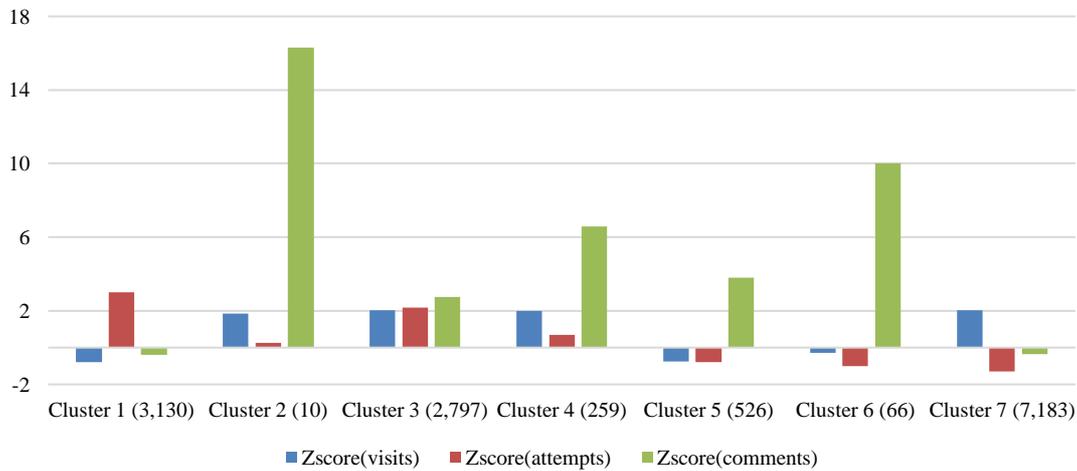

**Fig. 2.** Final cluster centers and sizes, when $k = 7$

Interestingly, Cluster 1 (3,130) and Cluster 7 (N = 7,183) had negative $Z_{comments}$ values thus containing, arguably, the "least social" students (large subpopulation, *i.e.* 73.82% of the whole population); yet this "low social" character could be associated with the highest $Z_{attempts}$ values (in case of Cluster 1) or the highest $Z_{visits}$ values (in case of Cluster 7). This suggests that whilst FutureLearn employs a social constructivist approach that encourages social interactions between students, [12, 24], there was a very large subpopulation (73.82%) reluctant to interact with peers but focusing on other activities, such as accessing learning materials (steps) and taking assessments (tests). This also suggests that being "less social" or "more social" does not necessarily predict how students were engaged in other activities. Additionally, even though they might not learn via direct social interaction with peers, they could still benefit from reading peers' discussions (comments).

On the contrary, although Cluster 5 (N = 526) and Cluster 6 (N = 66) had relatively high $Z_{comments}$, both their $Z_{visits}$ and $Z_{attempts}$ values were below zero. This indicates that there was a subpopulation (4.24%) that were focused on social interactions yet might not spend much time in other activities. The students allocated in these clusters might be the "contributors" of the MOOCs since they tended to share their thoughts which could be useful for other students.

## 5. Comparative Analysis

We next conducted comparative analysis based on the clustering result, as articulated above in section 4, to answer the second research question, ***RQ2: Are there behavioural and demographical patterns within subpopulations?***

First, we compared the three clustering variables between these seven clusters using boxplots. For *visits*, as shown in Fig. 3, Cluster 4 had the greatest mean and median, followed by Cluster 3, then Cluster 2. Cluster 1, Cluster 5, Cluster 6 and Cluster 7 had a lower median, yet Cluster 6's is much greater. This shows that, on average, students in Cluster 4 visited the largest number of *step*s; whilst students in clusters 5, 6, 7, and especially Cluster 1 visited the least number of *step*s. This may be because the students in these clusters dropped out from the course earlier. The box plots for Cluster 2 and Cluster 6 are much taller than those of the other five clusters, indicating the number of *step*s visited by students allocated in Cluster 2 and Cluster 6 was highly variable, indicating the variable *visits* is not useful in differentiating these two clusters.

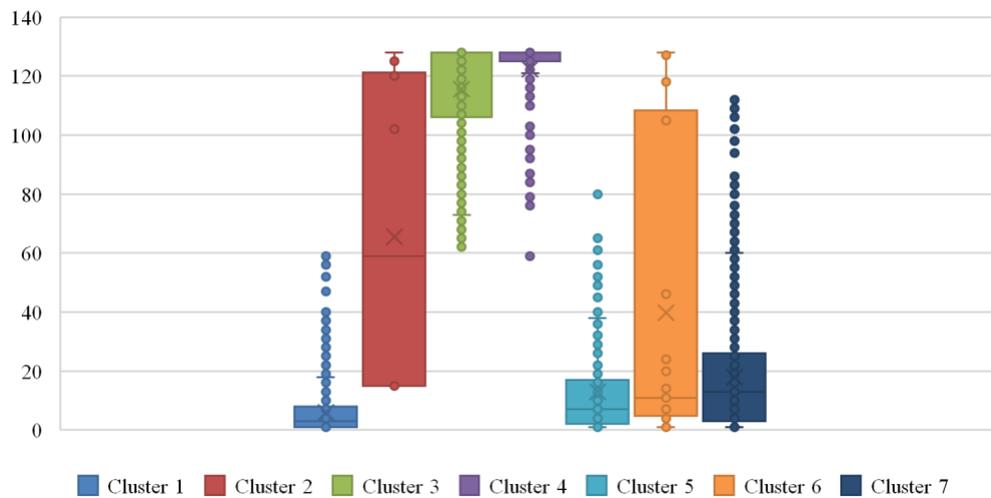

**Fig. 3.** Comparison of *visits* between clusters

For *attempts*, as shown in Fig. 4, Cluster 1, Cluster 3 and Cluster 4 had similar high mean and median. Cluster 7 had the lowest mean and median. This suggests that students in clusters 1, 3 and 4 attempted more times to answer questions in tests; whilst the students allocated in Cluster 7 were not interested in doing so. The heights of the box plots show a great variability for *attempts* of students in Cluster 5 and Cluster 6, indicating the variable *attempts* is not useful in differentiating these two clusters.

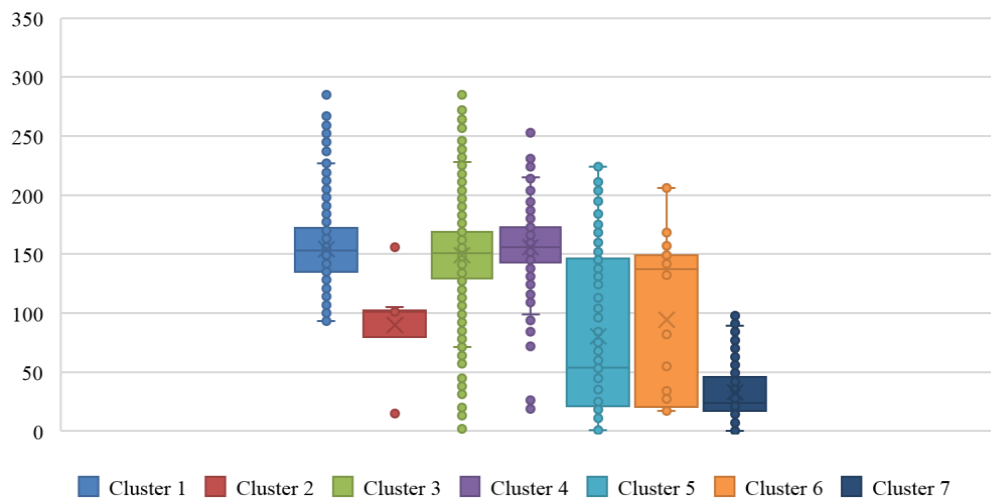

**Fig. 4.** Comparison of *attempts* between clusters

Regarding *comments*, as shown in Fig. 5, Cluster 2 had the highest mean and median, indicating students allocated in this cluster posted the greatest number of comments. Cluster 6 had the second greatest mean and median, followed by clusters 4 and 5. Interestingly, clusters 1, 3 and 7 had very low, close to zero, mean and median values, suggesting that students from these clusters tented not to post comments (or interact with peers or participate in discussions). The box plots representing *comments* are much shorter than those representing *visits* and *attempts*, which indicates students within each cluster had very similar commenting behavior, and that the variable *comments* is a distinguishing variable for all clusters.

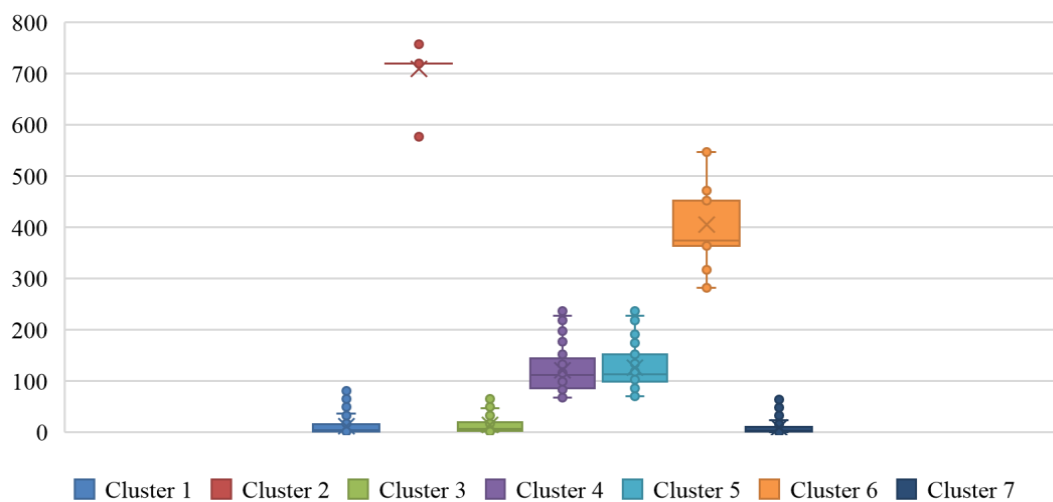

**Fig. 5.** Comparison of *comments* between clusters

To examine to what extend the clusters differ from each other, we conducted nonparametric Kruskal-Wallis H tests, for the three independent variables, *i.e. visits*, *attempts* and *comments*. The result suggests statistically significant differences between clusters: (1) *visits* ($\chi^2$ (2) = 7,931.41, $p < .001$), (2) *attempts* ($\chi^2$ (2) = 9,776.75, $p < .001$) and (3) *comments* ($\chi^2$ (2) = 2,772.40, $p < .001$).

**Table 3.** Mann-Whitney U test result – Asymp. Sig. (2-tailed) – *visits, attempts* and *comments*

| | *versus* | **Cluster 2** | **Cluster 3** | **Cluster 4** | **Cluster 5** | **Cluster 6** | **Cluster 7** |
|---|---|---|---|---|---|---|---|
| **Visits** | Cluster 1 | *< .001* | *< .001* | *< .001* | *< .001* | *< .001* | *< .001* |
| | Cluster 2 | | *< .001* | *< .001* | *< .001* | *0.042* | *0.003* |
| | Cluster 3 | | | *< .001* | *< .001* | *< .001* | *< .001* |
| | Cluster 4 | | | | *< .001* | *< .001* | *< .001* |
| | Cluster 5 | | | | | *0.001* | *< .001* |
| | Cluster 6 | | | | | | **.163** |
| **Attempts** | *versus* | **Cluster 2** | **Cluster 3** | **Cluster 4** | **Cluster 5** | **Cluster 6** | **Cluster 7** |
| | Cluster 1 | *< .001* | *< .001* | *0.047* | *< . 001* | *< .001* | *< .001* |
| | Cluster 2 | | *< .001* | *< .001* | **.674** | **.450** | *< . 001* |
| | Cluster 3 | | | *< .001* | *< .001* | *< .001* | *< .001* |
| | Cluster 4 | | | | *< .001* | *< .001* | *< .001* |
| | Cluster 5 | | | | | **.193** | *< .001* |
| | Cluster 6 | | | | | | *< .001* |
| **Comments** | *versus* | **Cluster 2** | **Cluster 3** | **Cluster 4** | **Cluster 5** | **Cluster 6** | **Cluster 7** |
| | Cluster 1 | *< .001* | *< .001* | *< .001* | *< .001* | *< .001* | *< .001* |
| | Cluster 2 | | *< .001* | *< .001* | *< .001* | *< .001* | *< .001* |
| | Cluster 3 | | | *< .001* | *< .001* | *< .001* | *< .001* |
| | Cluster 4 | | | | *< .001* | *< .001* | *< .001* |
| | Cluster 5 | | | | | *< .001* | *< .001* |
| | Cluster 6 | | | | | | *< .001* |

A Mann-Whitney U test for pairwise comparisons (Table 3) revealed clusters which, at significance level of .05, do not differ significantly: in *visits* Cluster 6 and Cluster 7 ($U$= 213,501.5), and, in *attempts*, Cluster 2 and Cluster 5 ($U$ = 2,426); Cluster 2 and Cluster 6 ($U$ = 281); Cluster 5 and Cluster 6 ($U$ = 15,654). Thus, the commenting behavior is most relevant for the clustering.

We further explored behavioral patterns of subpopulations by examining the following three indicators across our seven clusters: (1) **completion rate**: the number of *steps* a student claimed completion (by clicking the button "Mark as complete" on *step* pages), out of the number of distinct *steps* a student visited; (2) **correct answers rate**: the number of questions a student correctly answered, out of the number of questions a student answered in total; and (3) **reply rate**: the percentage of comments posted by a student that received replies.

We calculated a series of descriptive statistics to facilitate comparisons (see Table 4). Cluster 4 had the highest average *completion rate* (M = 97.56%, SD = 5.64%) followed by Cluster 3 (M = 95.95%, SD = 9.54%), while Cluster 1 had the lowest average *completion rate* (M = 37.21%, SD = 38.41%). For *correct answers rate*, all clusters did relatively well (>67.41%). Students from Cluster 5 (M = 90.53%, SD = 84.48%) and Cluster 7 (89.58%, 79.93%) performed the best, while students from Cluster 1 (M = 67.41%, SD = 12.71%) and Cluster 6 (M = 68.78%, SD = 14.17%) the worst. For *reply rate*, on average, Cluster 2 (M = 69.75%, SD = 18.15%) and Cluster 1 (M = 55.47%, SD = 24.46%) had the highest, and Cluster 4 (M = 29.65%, SD = 14.87%) and Cluster 3 (M = 34.65%, SD = 23.35%) the lowest. Interestingly, Cluster 4 had the highest *completion rate* yet lowest *reply rate*. This suggests that having better achievement in some aspect did not guarantee better achievement in all aspects. Kruskal-Wallis H tests showed statistically significant differences in achievement between the seven clusters. However, only some pairwise comparisons (Manny-Whitney U test) showed significant differences: in *completion rate*, Cluster 2 did not significantly differ from Cluster 6 (U = 257.5, p = 0.262) or Cluster 7 (U = 26,997.5, p = 0.171); in *correct answers rate*, Cluster 4 did not significantly differ from clusters 1, 3, 5 and 7; Cluster 1 did not significantly differ from clusters 3, 4 and 5; Cluster 3 did not significantly differ from clusters 1, 4 and 7; Cluster 5 did not significantly differ from clusters 1, 4 and 7; Cluster 2 did not significantly differ from Cluster 7; Cluster 6 did not significantly differ from Cluster 7; in *reply rate*: Cluster 1 did not significantly differ from clusters 3, 4 and 7; Cluster 2 did not significantly differ from clusters 3 and 4; Cluster 3 did not significantly differ from clusters 6 and 7; and Cluster 4 did not significantly differ from clusters 6 and 7. This shows, e.g., that the *correct answers rate* in Cluster 5 is not only the highest, but significantly so, even against the main competitor, *i.e.* Cluster 7.

**Table 4.** Students achievement in the course

|  |  | Cluster 1 | Cluster 2 | Cluster 3 | Cluster 4 | Cluster 5 | Cluster 6 | Cluster 7 |
|---|---|---|---|---|---|---|---|---|
| **Completion rate** | Mean | **37.21%** | 83.25% | 95.94% | **97.56%** | 50.49% | 62.56% | 62.40% |
|  | SD | 38.41% | 22.63% | 9.54% | 5.64% | 41.23% | 40.59% | 39.90% |
| **Correct answers rate** | Mean | **67.41%** | 73.47% | 75.42% | 70.70% | **90.53%** | 68.78% | 89.58% |
|  | SD | 12.71% | 7.71% | 62.01% | 12.01% | 84.48% | 14.17% | 79.93% |
| **Reply rate** | Mean | 55.47% | **69.75%** | 34.65% | **29.65%** | 40.28% | 48.33% | 47.24% |
|  | SD | 24.46% | 18.15% | 23.35% | 14.87% | 22.08% | 17.36% | 23.67% |

Next, we explored how clusters differed demographically from each other. Students' demographic data was collected using the pre-course survey. Overall, out of those 13,971 students, only 2,237 (16.01%) answered the question about their sex (515 male, 1,715 female, 1 "nonbinary", and 6 "other"). As "nonbinary" and "other" were very underrepresented (0.31%), to simplify the procedure, we considered only two categories: male and female. The overall sex ratio, *i.e.* the number of males per 100 females, was 30.03, very biased towards the female sex and much lower than human species' natural sex ratio at birth of 105 [25].

Fig. 6 compares the sex ratio between clusters. We can observe, while most of the clusters had similar sex ratio as that of the overall MOOC, *i.e.* 30.03, Cluster 2's sex ratio was the lowest, *i.e.* no male, and Cluster 6's sex ratio was the highest, *i.e.* 50. Taking these two "extreme" clusters for a further comparison, we found that Cluster 2 had both the highest $Z_{comments}$ (see Fig. 2) and the highest *reply rate* (see Table 4), whilst for Cluster 6, although it had the second highest $Z_{comments}$ (see Fig. 2), its *reply rate* (see Table 4) was much lower than that of Cluster 2. This is very interesting: females tended to post more comments, and their comments tended to attract more replies. Nevertheless, the type of comments, e.g. descriptive, debatable, challenging, encouraging, meaningful, just to name a few, might affect how likely their comments might receive replies, which is worth to further investigate.

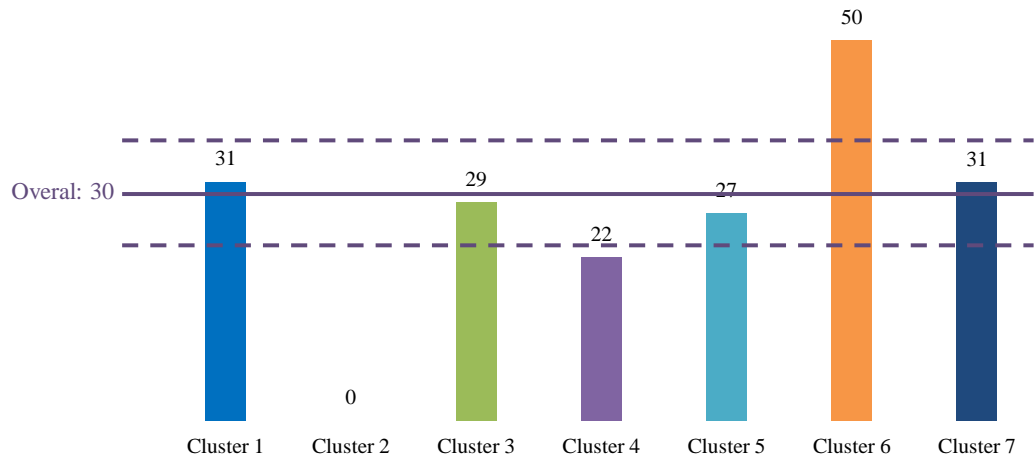

**Fig. 6.** Sex ratio (females/males) between seven clusters

Similar to the sex question in the pre-course survey, only 2,187 (15.65%) students answered the question about their age. These 2,187 responses included 31 (1.42%) as "<18", 205 (9.37%) as "18-25", 338 (15.45%) as "26-35", 276 (12.62%) as "36-45", 340 (15.55%) as "46-55", 464 (21.22%) as "56-65", and 533 (24.37%) as ">65". Interestingly, overall, older students occupied the largest portion of the population. One possible interpretation is that the MOOC investigated was humanities-themed, which might be more appealing to the subpopulation of older students, consistent with prior research [18]. Fig. 7 shows how the distribution of the age range varies between clusters: Cluster 4 has more older students; Cluster 5 has more younger students; the proportion of age ranges in clusters 1 and 7 were more even than in other clusters.

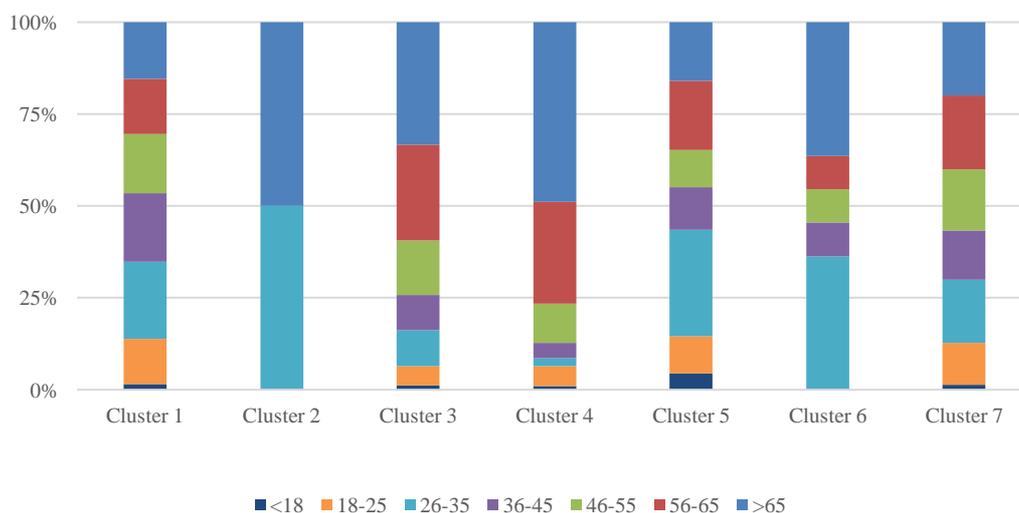

**Fig. 7.** Age range proportions within and between clusters

Thus, to *semantically analyze* our clusters, we can interpret them as follows:

- **Cluster 1** is a medium-sized cluster of 'quizzers', balanced in age, who try out many questions but can't really answer them (lowest significant *correct answers rate*), and who don't bother about completing (lowest *completion rate*). They would clearly **need intervention**, although motivating them only is not sufficient: they are trying but failing. They would need more simplified material and guided towards simpler questions.

- **Cluster 2** represents a very small number of extremely sociable (most *comments*), influential (highest *reply rate*) very young (18-25) or very old (over 65) females, who tend to complete most of the course but are average in terms of *correct answers rate*. They don't need immediate intervention, although help with their answers might be appropriate. It is clear that they enjoy the course.

- **Cluster 3** is a medium-sized cluster of students over-36-year-old, who have an excellent (second highest) *visit rate* and *completion rate*, who don't interact socially (almost no *comment*, second lowest *reply rate*), high number of *attempts* but medium *correct answers rate*. They don't need immediate intervention, although guiding them towards more social interaction, in terms of learning from other students and perhaps increasing the quality of their answers by this interaction might be appropriate.

- **Cluster 4** is a small cluster of older students with highest *visit rate* and *completion rate*, with some social interaction but no influence (lowest *reply rate*), highest number of *attempts* but medium *correct answers rate*. They also don't need immediate intervention, but, whilst they are more social than those in Cluster 3, they could still benefit more from learning from others.

- **Cluster 5** represents a quite small number of young, moderately sociable students with moderate influence on others, who answer a varying number of questions significantly well, but have the low number of *visits* and *completion rate*, *i.e.* about half of the course. These are students who **need intervention**. They may be either very busy, in which case rescheduling the remainder of the course might be appropriate, or, more concerning, then they might be bored with the learning material: they would potentially benefit from added challenges, to keep them participating in the course.

- **Cluster 6** is a very small number of highly sociable students with some influence on others, with a medium *completion rate* and *correct answers rate*. They don't need immediate intervention, although allowing them access to simpler learning material could increase their *correct answers rate* and participation in the course.

- **Cluster 7** is the largest group by far (more than double in size compared to the next clusters in size), with relatively even age distribution, who complete more than half of the *steps* they have visited, yet in a relatively low number of *visits*, but they don't do much else: the lowest number of *attempts* – although the questions they do answer have excellent *correct answers rate*; almost no comment – although for those who do comment, they get almost 50% *reply rate* to them. These students **need support**, as they seem demotivated, and removed socially – they need to be reminded that they should interact with other students and take tests, and this may convince them to stay in the course longer.

Surprisingly, clusters with the most completers (clusters 2, 3, 4) are not the ones with best *correct answers rate*. Moreover, as a majority, completers have the least influence over their fellow students (two with the lowest *reply rate* being Cluster 3, Cluster 4, which represent the majority of completers).

## 6.    Conclusions

In this study, we have identified three influential parameters, namely *visits*, *attempts,* and *comments*, which are independent enough to allow clustering students in a MOOC. Using the k-means algorithm and the "elbow method", we found 7 strong and stable subpopulations (clusters). We profiled these subpopulations, by comparing various behavioral and demographical patterns. Our method enabled the comparative analysis of the differences between subpopulations and possible interpretation of those differences.

This study contributes to a more in-depth understanding of how students are engaged in a MOOC, where student population can be extremely diverse, and this diversity can be extremely influential in how students behave and achieve. The insights found in the study can serve as indications to meet the diversity of behavioral and demographical patterns of student subpopulations in the "MOOCs context", which can guide the design of adaptive strategies that allow a better learning experience in MOOCs. Future research should focus on transforming behavioral and demographical patterns into meaningful predictors and intervenors for better adaptation and personalization in support of the heterogeneity and massiveness of MOOC students.